\documentclass[conference]{IEEEtran}
\usepackage{cite}
\usepackage{parskip}
\usepackage{amsmath}
\usepackage{hyperref}
\usepackage{amsfonts}
\usepackage{graphicx}

\usepackage{xcolor}

\begin{document}
\title{\vspace{-.8em}A Partial Break of the \emph{Honeypots Defense} \\ to Catch Adversarial Attacks}
\author{\IEEEauthorblockN{Nicholas Carlini (\emph{Google Brain})}}
\maketitle

\begin{abstract}
  A recent defense proposes to inject ``honeypots'' into neural
  networks in order to detect adversarial attacks.
  We break the baseline version of this defense by reducing the detection
  true positive rate to 0\%, and the detection AUC to 0.02, maintaining
  the original distortion bounds.
  The authors of the original paper have amended the defense in their
  CCS'20 paper to mitigate this attacks.
  To aid further research, we release the
  complete 2.5 hour  
  keystroke-by-keystroke  screen recording of our attack process at \\
  \url{https://nicholas.carlini.com/code/ccs_honeypot_break}.
\end{abstract}

\pagestyle{empty}
\thispagestyle{empty}

\section{Introduction}

Shan \emph{et al.} \cite{shan2019using} (CCS'20) recently proposed a honeypots-based
defense against adversarial examples.
This defense injects a backdoor into a neural network during training,
and then shows that
adversarial examples generated on this classifier
share similar activation patterns to backdoored inputs---and can
therefore be detected with near-perfect accuracy.

The authors of this paper provided us with early access to an
implementation of this defense.
We find that the baseline version of this defense is completely
ineffective.
We reduce the AUC to below $0.02$ (random guessing gives $0.50$), for
a true positive of $0\%$ at a false positive rate of $10\%$.
In response, the authors have amended the defense introducing
additional randomness and layers that mitigate this attack.
This short paper analyzes the baseline version of the defense.

\section{Attacking the Honeypot Defense}

We assume familiarity with prior work on adversarial examples \cite{szegedy2013intriguing},
and breaking adversarial examples detectors \cite{carlini2017adversarial}.
We use $f(x)$ to denote a trained neural network evaluated on input image $x$.
An adversarial example is an input $x'$ so that $\lVert x-x' \rVert$
is small (under some $\ell_p$ norm) but $f(x) \ne f(x')$.

\emph{The Honeypot Defense}
injects a backdoor perturbation
$\Delta$ during the neural network training process so that for all inputs $x$, the classifier will consistently and predictably misclassify $f(x + \Delta)$.
As a result of this backdoor, standard methods
to generate adversarial examples will create examples $x'$ that have
``characteristics'' of the backdoored inputs.

These characteristics are formalized by comparing the cosine similarity
between the hidden vectors $h(x')$ and the average backdoored hidden vector
$\phi = \mathbb{E}_{x \in \mathcal{X}}(h(x+\Delta))$.
That is, for a given input $x$, the defense computes
\[\text{sim}(h(x), \phi) = \frac{h(x) \cdot \phi}{\lVert h(x)\rVert \,\lVert\phi\rVert}\]
and rejects an input $x$ as adversarial if $\text{sim}(h(x), \phi) > \tau$.

\emph{Threat Model.}
This defense argues robustness under the $\ell_\infty$ norm threat
model ($\epsilon=8/255$) for both (a) a full white-box threat model, and
(b) a limited white-box threat model where the adversary has
access to the trained model $f_\theta$ but not the signature $\phi$.
The defense reports a $0.97$ minimum AUC across all prior
attacks, and claims a $0.76$ AUC against the strongest
adaptive attacks that allow a $6.3\times$ larger distortion bound.

\subsection{Initial White-Box Attack: Reducing AUC to 0.46}
Following recent advice \cite{tramer2020adaptive}, we design a loss function to be as simple as
possible to make it easy to diagnose difficulties in
optimization.
Thus, we adopt the most common attack technique, and minimize a weighted sum
of the misclassification loss and the detection-evading loss:
\begin{equation}
  \label{eqn}
  \mathop{\text{arg max}}_{\delta : \lVert \delta \rVert_\infty \le \epsilon}
  \mathcal{L}_{\text{xe}}\bigg(f(x+\delta),y\bigg) - \lambda \cdot \mathcal{L}_\text{d}\bigg(h(x+\delta),\phi\bigg)
\end{equation}
where $\mathcal{L}_{\text{xe}}$ is standard cross-entropy loss,
$\mathcal{L}_{\text{d}} = \text{sim}(h(x), \phi)$ is the loss of the detector
(defined on the signature $\phi$ for the hidden vector $h(\cdot)$),
and $\lambda$ is a hyperparameter controlling the relative importance
of the two terms\footnote{We set $\lambda=8$ for our attacks.
  Manual binary search determined that $\lambda=5$ was too small
  and $\lambda=10$ was too large. Setting $\lambda=8$ was just right.}.
We directly minimize this loss function with $100$ iterations of
$\ell_\infty$-regularized gradient descent with a step size of $0.1$.

This form of loss function is not new: we used it extensively in prior work \cite{carlini2017adversarial},
and the honeypot defense paper used it to
perform its own adaptive attack \cite{shan2019using}.
Nevertheless, the attack is effective at reducing the defense
AUC to $0.46$---below the threshold of $0.5$ corresponding to
random guessing.
We are unable to explain why our attack succeeded when the authors
attempt at this exact formulation failed.

\subsection{Improved White-Box Attack: Reducing AUC to 0.02}
The above loss formulation has a weakness: an optimal attack
method should satisfy three constraints simultaneously.
The final generated adversarial example:
\begin{itemize}
\item should introduce a sufficiently small perturbation;
\item should be misclassified as a particular target class; and,
\item should not be detected as adversarial by the detector.
\end{itemize}

Solving Equation \ref{eqn} guarantees that the perturbation is
bounded correctly (because of the hard constraint),
but does not guarantee the other two properties.
When minimizing Equation \ref{eqn}, we might
over-optimize the cross-entropy
loss at the expense of the detection loss (if $\lambda$ is
too small) or instead that we might might over-optimize the detection loss
(if $\lambda$ is too big).
Instead, we would like to ensure that whenever the input
already adversarial, all available distortion ``budget''
goes into fooling the detector (and vice versa).

We thus consider an improved attack that alternates between
two gradient descent procedures.
As long as the input $x+\delta$ is misclassified, i.e., $f(x+\delta)\ne y$, we perform straightforward
gradient descent minimizing the detection loss:
\begin{equation}
  \delta \gets \text{proj}_{\lVert\delta\rVert\le\varepsilon} \big(\delta - \eta \cdot \nabla\mathcal{L}_\text{d}(h(x+\delta),\phi)\big)
\end{equation}
taking steps of size $\eta$ and ensuring the perturbation remains
bounded within the $\ell_\infty$ box with norm $\epsilon$.

Alternatively, if instead $f(x+\delta)=y$, then we minimize the
cross-entropy loss.
As a first attempt we update with
\begin{equation}
  \label{next}
  \delta \gets \text{proj}_{\lVert\delta\rVert\le\varepsilon} \big(\delta + \eta \cdot \nabla\mathcal{L}_\text{xe}(f(x+\delta),y)\big).
\end{equation}
By doing this, we can ensure that every gradient descent step is
helpful: when $x+\delta$ is misclassified we take steps to reduce
the likelihood it is detected; when $x+\delta$ is not misclassified
we take steps to increase the cross entropy loss.

This has one drawback: often these two steps point in
opposite directions.
Progress is then slow, with each step ``undoing'' the progress
made in the prior step.
To alleviate this, whenever we take steps to make the input more adversarial,
we ensure that doing so does not also make the input more detectable.
This is achieved by ensuring that all cross-entropy steps are orthogonal to
the detection gradient direction.
Formally, let 
\begin{align}
  \label{eqn1}
  g_x & = \nabla\mathcal{L}_\text{xe}(f(x+\delta),y) \\
  g_d & = \nabla\mathcal{L}_\text{d}(h(x+\delta),\phi)
\end{align}
then we replace Equation \ref{next} with
\begin{equation}
  \label{eqn2}
  \delta \gets \text{proj}_{\lVert\delta\rVert\le\varepsilon} \bigg(\delta + \eta \cdot \big(g_x - g_d \frac{g_d \cdot g_x}{\lVert g_d \rVert \lVert g_x \rVert} \big)\bigg).
\end{equation}

These two approaches are identical when allowed a sufficient number
of iterations of gradient descent.
However, it is easy to see why this procedure is more efficient
for a limited number of gradient descent steps: for sufficiently
small step sizes $\eta$, the update rule in Equation \ref{eqn2}
is guaranteed to be orthogonal the gradient direction from Equation \ref{eqn1}.
Therefore, we never make negative progress on steps in this direction.
This improved attack reduces the classifier AUC to $0.02$.

\subsection{Attacking without Signature Knowledge}
The defense also claims robustness against an adversary who
is not aware of the signature $\phi$.
Unfortunately, the defense is also broken under this threat model.
Because of the intuition of the defense---that typical adversarial
examples will have a signature similar to $\phi$---it is possible to
estimate it through
\[\tilde\phi = \mathbb{E}_{x \in \mathcal{X}}(h(\mathcal{A}(x))\]
where $\mathcal{A}(x)$ generates an adversarial example on input $x$.
Then we run exactly the prior attack substituting $\tilde\phi$ for $\phi$.

More generally, consider an adversary who computes two adversarial
examples $x'$ and $x''$ for a given input $x$ such that $h(x') \cdot h(x'') = 0$.
Then by randomly returning one of these inputs as the result of
$\mathcal{A}(x)$,
is will be definitionally impossible for the classifier to obtain
greater than a $50\%$ true positive rate.

\subsection{Mitigating this Attack}

The honeypot defense authors have mitigated this attack in the final version of their paper.
We do not analyze the robustness of this modified scheme, and
refer the reader to the updated paper for details on how the
scheme has been modified.
It is an interesting and open question to study if the improved defense
could be evaded with a stronger attack.

\section{Discussion}

The attacks presented above are simple modifications
of well-known methods, and apply gradient descent to a well-crafted loss function.
This phenomenon is not new---an appropriate implementation of gradient descent
has sufficed for breaking many defenses published over the last several years \cite{tramer2020adaptive}.

Although we should not require that published defenses
be perfect and resist all attack,
we should hope that attacks on published defenses
require novel attack approaches.
Even when defenses can be broken, if they require sophisticated
attacks then they can be extremely valuable in order
to help better understand what are and are not fundamental properties
of adversarial examples.
However, when breaks amount to ``apply gradient descent'', 
there are few generalizable lessons other than that
one particular idea does not work.

In order to provide more perspective,
we recorded our 2.5 hour attack, keystroke-by-keystroke,
to document the steps we follow.
(This two and a half hours goes from first
inspecting the code to the final break, and is
not an atypical
amount of time; attacks in \cite{tramer2020adaptive} took similarly long.)
We hope this additional artifact might provide useful for
developing improved procedures for assessing performance of studied defenses: \\ 
\url{https://nicholas.carlini.com/code/ccs_honeypot_break}

\section*{Acknowledgements}
We are grateful to Shawn Shan and Ben Zhao for providing us 
code and discussing their defense, and
Aleksander Madry for comments on an early draft of this paper.

{\footnotesize
\bibliographystyle{IEEEtranS}
\bibliography{paper}

\begin{thebibliography}{1}
\providecommand{\url}[1]{#1}
\csname url@samestyle\endcsname
\providecommand{\newblock}{\relax}
\providecommand{\bibinfo}[2]{#2}
\providecommand{\BIBentrySTDinterwordspacing}{\spaceskip=0pt\relax}
\providecommand{\BIBentryALTinterwordstretchfactor}{4}
\providecommand{\BIBentryALTinterwordspacing}{\spaceskip=\fontdimen2\font plus
\BIBentryALTinterwordstretchfactor\fontdimen3\font minus
  \fontdimen4\font\relax}
\providecommand{\BIBforeignlanguage}[2]{{%
\expandafter\ifx\csname l@#1\endcsname\relax
\typeout{** WARNING: IEEEtranS.bst: No hyphenation pattern has been}%
\typeout{** loaded for the language `#1'. Using the pattern for}%
\typeout{** the default language instead.}%
\else
\language=\csname l@#1\endcsname
\fi
#2}}
\providecommand{\BIBdecl}{\relax}
\BIBdecl

\bibitem{carlini2017adversarial}
N.~Carlini and D.~Wagner, ``Adversarial examples are not easily detected:
  Bypassing ten detection methods,'' \emph{AISec}, 2017.

\bibitem{shan2019using}
S.~Shan, E.~Wenger, B.~Wang, B.~Li, H.~Zheng, and B.~Y. Zhao, ``Using honeypots
  to catch adversarial attacks on neural networks,'' \emph{CCS}, 2020.

\bibitem{szegedy2013intriguing}
C.~Szegedy, W.~Zaremba, I.~Sutskever, J.~Bruna, D.~Erhan, I.~Goodfellow, and
  R.~Fergus, ``Intriguing properties of neural networks,'' 2014.

\bibitem{tramer2020adaptive}
F.~Tramer, N.~Carlini, W.~Brendel, and A.~Madry, ``On adaptive attacks to
  adversarial example defenses,'' \emph{arXiv preprint arXiv:2002.08347}, 2020.

\end{thebibliography}
}

\end{document}